# A substantial increase of Curie temperature in a new type of diluted magnetic semiconductors via effects of chemical pressure

Shuang Yu[1,2], Guoqiang Zhao[1,2], Yi Peng[1,3], Xiaohong Zhu[3], Xiancheng Wang[1,2], Jianfa Zhao[1,2], Lipeng Cao[1,2], Wenmin Li[1,2], Zhi Li[4,a)], Zheng Deng[1,2,b)], and Changqing Jin[1,2,5,c)]

[1] *Beijing National Laboratory for Condensed Matter Physics, and Institute of Physics, Chinese Academy of Sciences, Beijing 100190, China*
[2] *School of Physics, University of Chinese Academy of Sciences, Beijing 100190, China*
[3] *Department of Materials Science & Engineering, Sichuan University, Chengdu, China*
[4] *School of Materials Science and Engineering, Nanjing University of Science and Technology, 210094 Nanjing, China*
[5] *Materials Research Lab at Songshan Lake, 523808 Dongguan, China*

a)Electronic mail: zhili@njust.edu.cn

b)Electronic mail: dengzheng@iphy.ac.cn

c)Electronic mail: Jin@iphy.ac.cn

Chemical pressure is an effective method to tune physical properties, particularly for diluted magnetic semiconductors (DMS) of which ferromagnetic ordering is mediated by charge carriers. Via substitution of smaller Ca for larger Sr, we introduce chemical pressure on $(Sr,Na)(Cd,Mn)_2As_2$ to fabricate a new DMS material $(Ca,Na)(Cd,Mn)_2As_2$. Carriers and spins are introduced by substitutions of (Ca,Na) and (Cd,Mn) respectively. The unit cell volume reduces by 6.2% after complete substitution of Ca for Sr, suggesting a subsistent chemical pressure. Importantly the local geometry of $[Cd/MnAs_4]$ tetrahedron is optimized via chemical compression that increases the Mn-As hybridization leading to enhanced ferromagnetic interactions. As a result, the maximum Curie temperature ($T_C$) is increased by about 50% while the the maximum saturation moment increases by over 100% from $(Ca,Na)(Cd,Mn)_2As_2$ to $(Sr,Na)(Cd,Mn)_2As_2$. The chemical pressure estimated from the equation of state is equal to an external physical pressure of 3.6 GPa which could be obtained via strain-engineering in thin films.

The diluted magnetic semiconductors (DMS) have been investigated extensively as they offer an opportunity to control the ferromagnetic properties by changing carrier density. The advantage leads to potential applications in spintronic devices.[1-3] Generally the spin & charge doping are induced by one element doping such as Mn doping into (Ga,Mn)As leading to difficulty in tuning either conducting or magnetic properties.[4] Recently a series of new type of DMS materials with independent carrier and spin doping have been discovered to overcome aforementioned difficulty, e.g. $Li_{1+x}$(Zn,Mn)As termed "111" type or (Ba,K)(Zn,Mn)$_2$As$_2$ (BZA) termed "122" type. BZA holds the record of Curie temperature ($T_C$ = 230 K) among carrier-mediated ferromagnetic DMS.[4-8]

Given a DMS material, effective ways to modify $T_C$ can be achieved by increasing the carrier density using an applied electric field, photoexcitations or pressure.[4,9] Particularly, pressure is expected to increase both carrier concentration and Mn-As hybridization which result in an enhancement of ferromagnetic interactions in DMS materials.[10] On the other hand, internal chemical pressure and epitaxial strain which play a comparable role as external physical pressure, are widely used to modify physical properties in many functional materials. For instance, an equivalent increase in superconducting critical temperature in cuprate superconductors has been reported via relatively low pressures (4-6 GPa) induced by film strains.[11,12] Superconductivity in iron-based compound BaFe$_2$As$_2$ can be induced by moderate pressure (<6 GPa) and iso-valent chemical doping (BaFe$_2$As$_{2-x}$P$_x$) respectively.[13,14] Comparing to external physical pressure, internal chemical pressure which can be applied by iso-valent substitutions, does not require any specific devices (e.g. diamond anvil cell or piston cylinder cell). Nevertheless, chemical pressure-effects in DMS materials are rarely reported.

Previous studies of physical pressure-effects on "122" BZA only presented negative pressure-effect on $T_C$. The proposed reason is that physical pressure distorts [MnAs$_4$] tetrahedra and then reduces effective Mn-As hybridization which in turn damages ferromagnetic ordering.[15-17] In this work we generated chemical pressure by changing atom size on another group of DMS (Sr,Na)(Cd,Mn)$_2$As$_2$.[18] By replacement of Sr for Ca, (Ca,Na)(Cd,Mn)$_2$As$_2$ was synthesized as a new DMS material. From Sr- to Ca-compound, the unit cell volume decreases by 6.2% suggesting positive chemical pressure effect. It is found that local geometry of [MnAs$_4$] tetrahedron in (Ca,Na)(Cd,Mn)$_2$As$_2$ is optimized by chemical pressure. Consequently, a successful improvement of ferromagnetic ordering by chemical pressure has been observed: comparing to (Sr,Na)(Cd,Mn)$_2$As$_2$, both maximum Curie temperature and saturation moment in (Ca,Na)(Cd,Mn)$_2$As$_2$ are significantly enhanced.

Both CaCd$_2$As$_2$ and SrCd$_2$As$_2$ crystallize into hexagonal structure with *P-3m1* space group (No. 164) as shown in Figure 1(a). Powder X-ray diffraction patterns for samples show that all of the peaks can be well indexed into *P-3m1* space group as shown in Figure S1. The lattice constants were calculated by Rietveld refinement. Both of *a*- and *c*-axis shrink linearly with increasing Mn doping level in Figure 1(b) because Mn$^{2+}$ (0.66 Å) is smaller than Cd$^{2+}$ (0.78 Å), well following the Vegard Law, an evidence of successful (Cd,Mn) substitution. CaCd$_2$As$_2$ and SrCd$_2$As$_2$ are quasi 2D-materials where Ca/Sr ions layers and honeycomb-like Cd$_2$As$_2$ layers stack alternately along c axis.[19] Given lattice constants for SrCd$_2$As$_2$ (*a* ~ 4.4516 Å, *c* ~ 7.4221 Å, *V* ~ 127.4Å$^3$) and CaCd$_2$As$_2$ (*a* ~ 4.3909 Å, *c* ~ 7.1870 Å, *V* ~ 120.0 Å$^3$), chemical compression effect is visible in the latter, particularly along c-axis. Besides, two more principal deviations between $CaCd_2As_2$ and $SrCd_2As_2$ is the Cd/Mn-As bond lengths and As-Cd/Mn-As bond angles in $Cd_2As_2$ layers which will be discussed in more details.

Figure 2(a) shows temperature dependent of magnetization ($M(T)$) curves for $(Ca_{1-x}Na_x)(Cd_{1-y}Mn_y)_2As_2$ (x = 0.025, 0.05, 0.1, 0.15; y = 0.05, 0.15, 0.2) under field $H$ = 500 Oe. There is no obvious difference between zero field cooling (ZFC) and field cooling (FC) but clear ferromagnetic signatures are observed for all samples, i.e. sharp upturns with decreasing temperature. Above $T_C$, susceptibility is fitted with Curie-Weiss law (inset of Figure 2(a)), $(\chi-\chi_0)^{-1} = (T - \theta)/C$, where $\chi_0$ stands for a temperature-independent term and $\theta$ for paramagnetic temperature. Neither $T_C$ nor $\theta$ monotonically increases with increasing Mn or Na doping level (Figure 2(c)). Maximum $T_C$ ~ 19 K and $\theta$ ~ 22 K are obtained for x = 0.05, y = 0.15. The maximum $T_C$ of (Ca,Na)(Cd,Mn)$_2$As$_2$ is about 50% higher than that of (Sr,Na)(Cd,Mn)$_2$As$_2$ (the maximum $T_C$ ~ 13 K).[18] After reaching maximum $T_C$, ferromagnetic ordering is weakened by over-doped Na or Mn, similar to analogues (Sr,Na)(Zn,Mn)$_2$As$_2$ and (Sr,Na)(Cd,Mn)$_2$As$_2$.[18,20] Effective paramagnetic moments ($M_{eff}$) are calculated from the Curie constant C. For example $M_{eff}$ of $(Ca_{0.95}Na_{0.05})(Cd_{0.95}Mn_{0.05})_2As_2$ is close to $5\mu_B$/Mn. Ferromagnetic characteristics, which are spontaneous magnetization under very low fields, narrow but clear hysteresis loops, are also found in $M(H)$ curves as plotted in Figure 2(b). Coercive fields are smaller than 100 Oe. Saturation moments ($M_{sat}$) decrease with increasing Mn (Figure 2(d)), due to increased competition between antiferromagnetic and ferromagnetic interactions as reported in other DMS systems. Nevertheless, maximum $M_{sat}$ of (Ca,Na)(Cd,Mn)$_2$As$_2$ is significant larger than that of (Sr,Na)(Cd,Mn)$_2$As$_2$ (Maximum $M_{sat} < 1\mu_B$/Mn). The larger $M_{sat}$ indicates that more local spins on Mn are ferromagnetic ordered, consistent with higher $T_C$ in (Ca,Na)(Cd,Mn)$_2$As$_2$.[18]

Electrical transport measurements are shown in Figure 3. The temperature dependent resistivity ($\rho(T)$) for parent compound CaCd$_2$As$_2$ shows semiconducting behavior within temperature range of 2 – 300 K (Figure 3(a)). It is worth noting that the resistivity of CaCd$_2$As$_2$ is much smaller than SrCd$_2$As$_2$ ($\rho_{300K}$ ~ 1*10$^4$ Ω·mm and $\rho_{120K}$ ~ 1*10$^7$ Ω·mm).[18] It is consistent with the aforementioned scenario that shortened Cd/Mn-As bond lengths and optimized As-Zn/Mn-As bond angle within sub-layers enhance intra-sub-layer Cd/Mn-As hybridization and in turn benefit conduction. On the other hand, $\rho_{2K}$ of CaCd$_2$As$_2$ is 3 orders magnitude larger than all the Na-doped (Ca,Na)(Cd,Mn)$_2$As$_2$, indicating significantly increased carrier concentrations via Na doping. The scheme is further supported as shown in Figure 3(a) by the decrease of resistivity of (Ca$_{1-x}$Na$_x$)(Cd$_{0.85}$Mn$_{0.15}$)$_2$As$_2$ with increasing Na-doping level. In contrast, as shown in Figure 3(b), resistivity of (Ca$_{0.95}$Na$_{0.05}$)(Cd$_{1-y}$Mn$_y$)$_2$As$_2$ gradual increases with increasing Mn concentrations.

Figure 4(a) shows $\rho(T)$ curves for (Ca$_{0.95}$Na$_{0.05}$)(Cd$_{0.85}$Mn$_{0.15}$)$_2$As$_2$ under various fields. Negative magnetoresistance (MR = $\Delta\rho/\rho_0$ = ($\rho_H$-$\rho_0$)/$\rho_0$) is found below ~ 18 K consistent with $T_C$ from magnetization data. Above 18 K, positive MR emerges. The consistency indicates that the negative MR is related to ferromagnetic ordering. In Figure 4(b), MR doesn't saturate at $H$ = 7 T and $T$ = 2 K, where the spins are almost fully aligned according to $M(H)$ curve. In (Ga,Mn)As, and analogue (Sr,Na)(Cd,Mn)$_2$As$_2$ the unsaturated MR are explained with giant splitting of the valence band. In order to understand such behavior, the negative magnetoresistance dates at 2 K are fitted with following equation,[21-23]

$$\Delta \rho / \rho_H = \Delta \sigma / \sigma = kB^{1/2} = -n_v e^2 C_0 \rho (eB\hbar)^{1/2} / (2\pi^2 \hbar), \qquad (1)$$

where $C_0 \approx 0.605$, $e$ is the elemental charge, $\hbar$ is the reduced Planck constant, respectively, and $1/2 \leq n_v \leq 2$ depending on the number of hole sub-bands contributing to the charge transport. The best fitting to Eq. (1) gives $n_v = 0.62$, close to that of $(Sr,Na)(Cd,Mn)_2As_2$. The maximum MR is ~15% at $T = 2$ K and $H = 7$ T. It is larger than analogues $(Sr,Na)(Zn,Mn)_2As_2$ and $(Ca,Na)(Zn,Mn)_2As_2$ as well as $(Ba,K)(Zn,Mn)_2As_2$ which has a much higher $T$c.[7,24-25]

Considering key roles of local geometry of $[Zn/MnAs]_4$ tetrahedra to ferromagnetic interaction in BZA, we compare Cd/Mn-As bond lengths and As-Cd/Mn-As bond angles of $CaCd_2As_2$ and $SrCd_2As_2$ to seek microscopic insight into the origin of improved ferromagnetic ordering in $CaCd_2As_2$. For carrier-mediated ferromagnetism in DMS, itinerant carriers play an important role in ferromagnetic interaction.[26-35] Given the quasi 2D structural of $CaCd_2As_2$ and $SrCd_2As_2$, one can expect carriers are more itinerant along ab-plane than c-axis. If one takes a close look at $Cd_2As_2$ planes, it is easy to find two sub-layers within one CdAs plane (Figure 1 (a)). It is reasonable to assume that intra-sub-layer component is more important than inter-sub-layer one to modify carriers mobility within the $Cd_2As_2$ plane. With the same doping levels, sub-layer of $CaCd_2As_2$ has shorter Cd/Mn-As bond length and more optimal As-Cd/Mn-As bond angles than that of $SrCd_2As_2$. As shown in Figure 1 (c) the $(Ca_{0.95}Na_{0.05})(Cd_{0.95}Mn_{0.05})As_2$ has the average Cd/Mn-As bond length of 2.700 Å and the average As-Cd/Mn-As bond angle within sub-layers of 108.9° that is close to the ~109.47° for a no distorted ideal tetrahedron.[15] On the other hand in $(Sr_{0.95}Na_{0.05})(Cd_{0.95}Mn_{0.05})As_2$ the average Cd/Mn-As bond length is 2.712 Å and the average As-Cd/Mn-As bond angle is 113.6° that is apparently deviated from the ~109.47°. The shortened Cd/Mn-As bond length will definitely increase Mn-As hybridization. Additionally the ideal As-Cd/Mn-As bond angle will increase the overlap of Mn-As planar orbitals and guarantee the maximum strength of Mn-As hybridization hence benefits to increase the ferromagnetic interactions. Previous studies of physical pressure-effects on "122" BZA indicated that shortened Zn/Mn-As bond length and optimized As-Zn/Mn-As bond angle (~109.47° for a regular tetrahedron) will enhance Cd/Mn-As hybridization.[15] In short (Ca,Na)(Cd,Mn)$_2$As$_2$ have stronger intra-sub-layer Cd/Mn-As hybridization than that for (Sr,Na)(Cd,Mn)$_2$As$_2$. As a result, we found improved ferromagnetic ordering in (Ca,Na)(Cd,Mn)$_2$As$_2$.

We calculated the equation of state (EoS) equation with first-principles calculations with plane augmented-wave (PW) pseudopotential method implemented in VASP code[36] to build up relationship between cell volume and pressure ($P(V)$) of SrCd$_2$As$_2$. Based on $P(V)$ curve in Figure S2, we estimate that an external pressure of 3.6 GPa can reduce cell volume of SrCd$_2$As$_2$ to 120.0 Å$^3$ (volume of CaCd$_2$As$_2$ at ambient pressure). In thin films an equivalent high pressure state can be induced by epitaxial strain. According to previous studies, a compressive pressure of 3.6 GPa could be obtained by tolerant lattice mismatch in thin films.[11] Thus the improvement of $T_C$ via internal chemical pressure in bulk (Ca,Na)(Cd,Mn)$_2$As$_2$ could in principle be induced by epitaxial strain in thin film state.

In summary, we successfully synthesized a new type of DMS, (Ca,Na)(Cd,Mn)$_2$As$_2$. The carriers and spins are introduced via (Ca,Na) and (Cd, Mn) substitutions independently. The Curie temperature of (Ca,Na)(Cd,Mn)$_2$As$_2$ is 50% higher than that for (Sr,Na)(Cd,Mn)$_2$As$_2$ due to the effects of chemical pressure, and the saturation moments is also enhanced dramatically. The significant improvement of ferromagnetism in (Ca,Na)(Cd,Mn)$_2$As$_2$ indicate the prospect to search for high temperature diluted magnetic semiconductors via proper chemical pressure and strain-engineering in thin films.

This work was financially supported by National Key R&D Program of China (No. 2017YFB0405703), Ministry of Science and Technology of China (2018YFA03057001，2015CB921000) &National Natural Science Foundation of China through the research projects (11534016).

**Figures**

**Fig. 1**

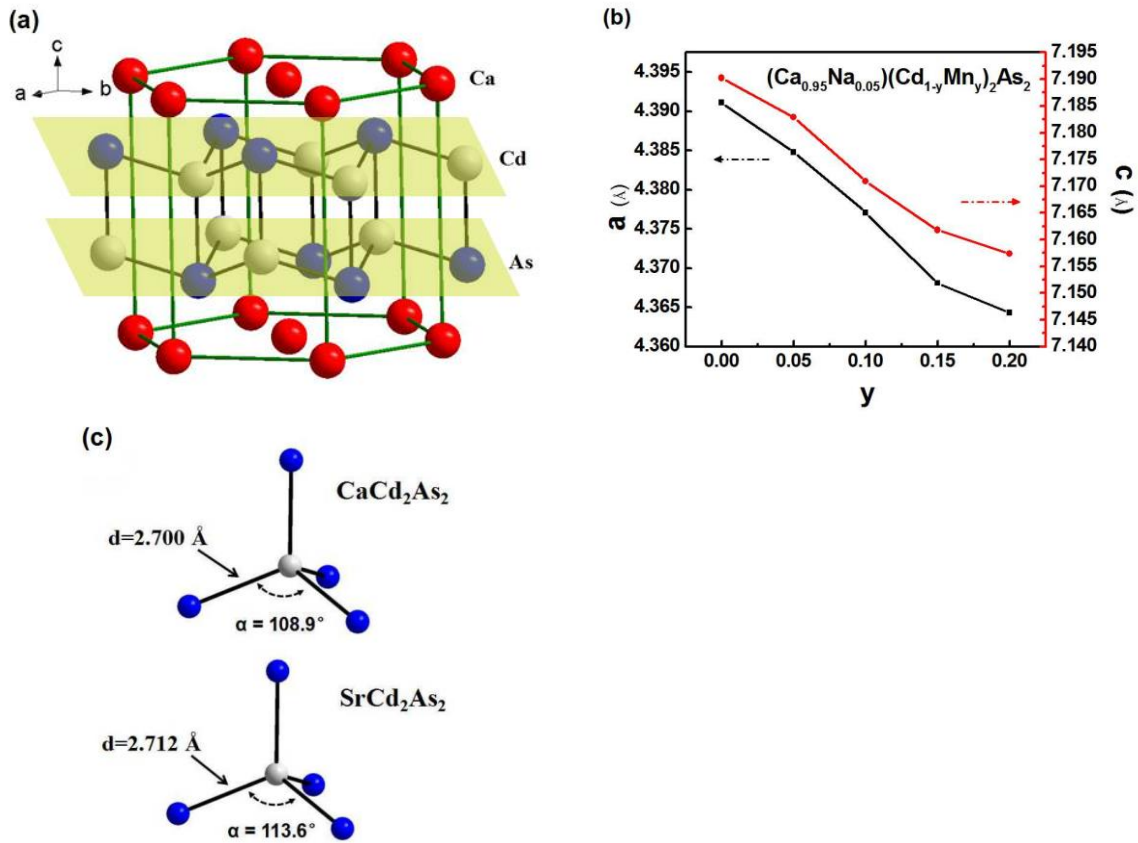

**Figure 1.** (a) Crystal structure of the parent phase, $CaCd_2As_2$. The CdAs sub-layers are highlighted with yellow parallelograms. (b) Lattice constants versus Mn doping levels. (c) [Cd/MnAs]$_4$ tertrahedra in $(Ca_{0.95}Na_{0.05})(Cd_{0.95}Mn_{0.05})As_2$ and $(Sr_{0.95}Na_{0.05})(Cd_{0.95}Mn_{0.05})As_2$. Marked bond length and bond angle are the ones within the CdAs sub-layers.

**Fig. 2**

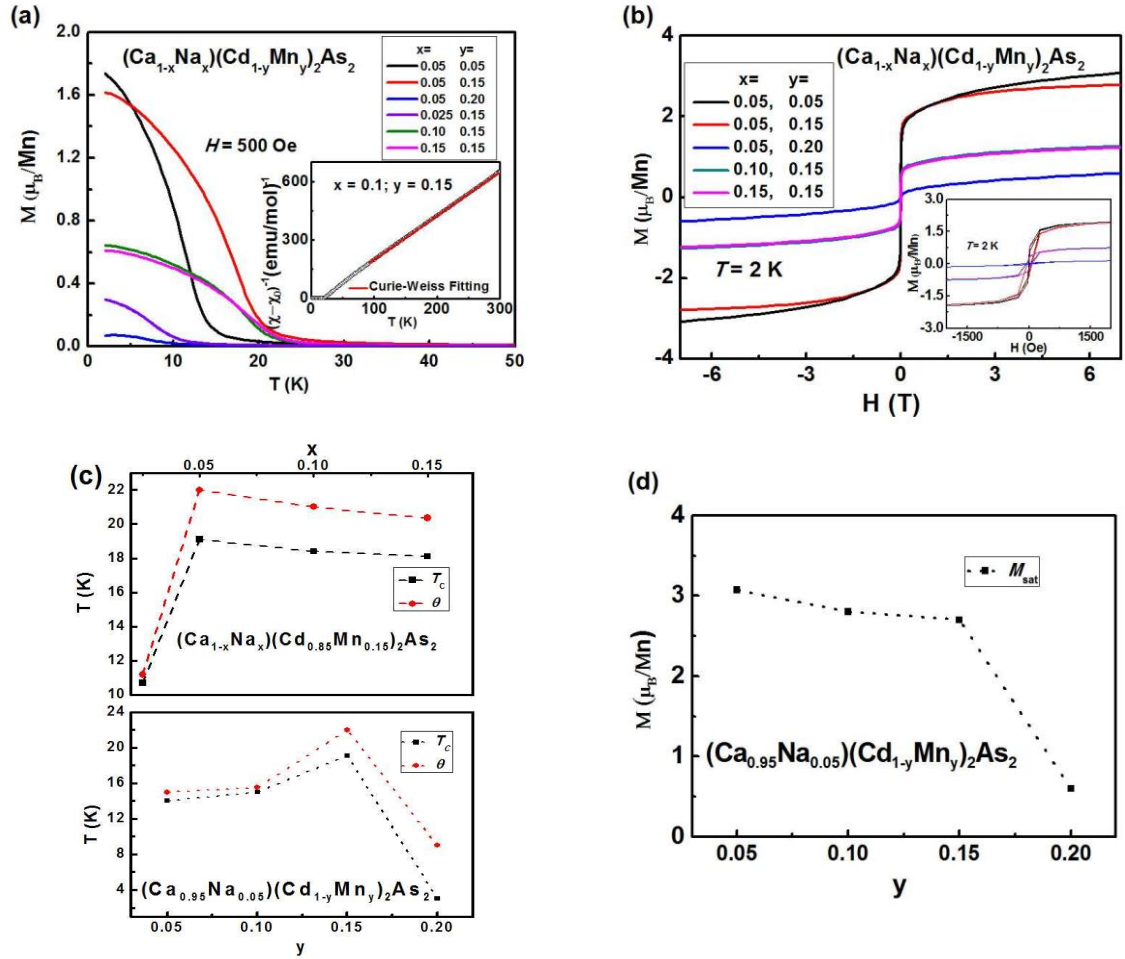

**Figure 2.** (a) $M(T)$ measured under $H$ =500 Oe of $(Ca_{1-x}Na_x)(Cd_{1-y}Mn_y)_2As_2$ (x = 0.025, 0.05, 0.1, 0.15; y = 0.05, 0.15, 0.2). (b) The hysteresis loops at 2 K for $(Ca_{1-x}Na_x)(Cd_{1-y}Mn_y)_2As_2$ (x = 0.025, 0.05, 0.1, 0.15; y = 0.05, 0.15, 0.2). (c) $T_C$ and $\theta$ versus Na- and Mn doping level. (d) $M_{sat}$ versus Mn doping level.

**Fig. 3**

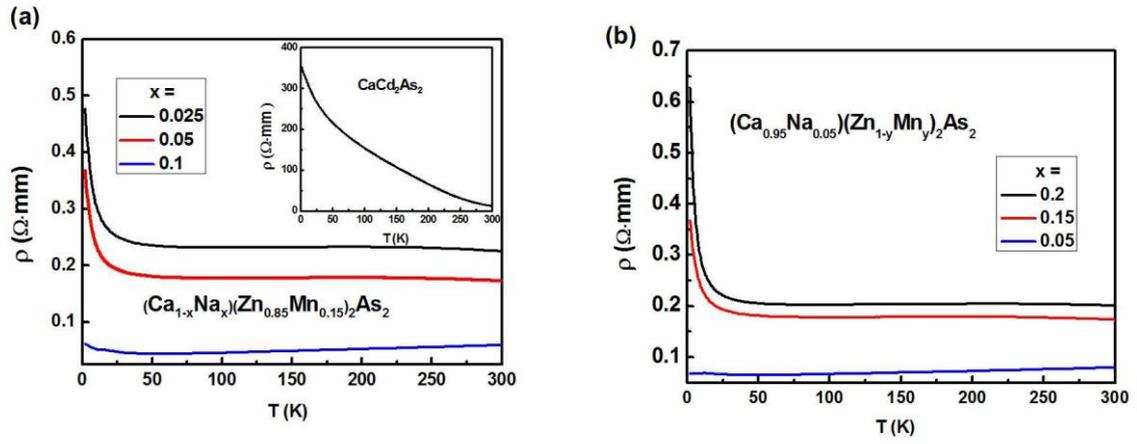

**Figure 3.** Temperature dependent resistivity curves of **(a)** $(Ca_{1-x}Na_x)(Cd_{0.85}Mn_{0.15})_2As_2$, (x = 0.025, 0.05, 0.1) and $CaCd_2As_2$ in the insert. **(b)** $(Ca_{0.95}Na_{0.05})(Cd_{1-y}Mn_y)_2As_2$ (y = 0.05, 0.15, 0.2).

**Fig. 4**

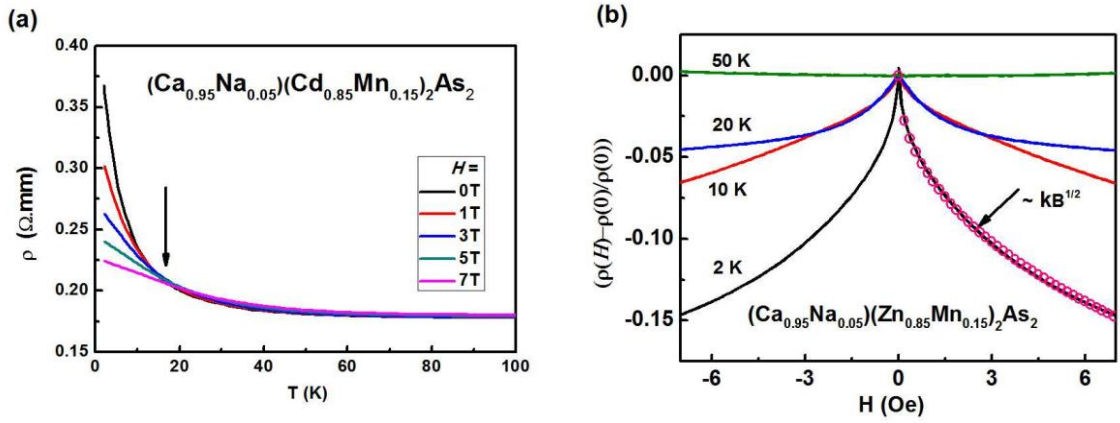

**Figure 4. (a)** $\rho(T)$ curves of $(Ca_{0.95}Na_{0.05})(Cd_{0.85}Mn_{0.15})_2As_2$ under various field. **(b)** Magnetoresistance curves of $(Ca_{0.95}Na_{0.05})(Cd_{0.85}Mn_{0.15})_2As_2$ measured in an external field up to 7 T at $T$ = 2, 10, 20, and 50 K, respectively. The pink circles show the fitting result according to Eq. (1).